\documentclass[conference]{IEEEtran}

\IEEEoverridecommandlockouts
\usepackage{balance}
\usepackage{cite}
\usepackage{amsmath,amssymb,amsfonts}
\usepackage{algorithmic}
\usepackage{graphicx}
\usepackage[usenames,dvipsnames,table]{xcolor}
\usepackage{textcomp}

\def\BibTeX{{\rm B\kern-.05em{\sc i\kern-.025em b}\kern-.08em
    T\kern-.1667em\lower.7ex\hbox{E}\kern-.125emX}}

\usepackage{xspace}

\newcommand*{\cf}{\textit{cf.}\@\xspace}
\newcommand*{\eg}{\textit{e.g.}\@\xspace}
\newcommand*{\ie}{\textit{i.e.}\@\xspace}
\makeatletter
\newcommand*{\etc}{%
    \@ifnextchar{.}%
        {etc}%
        {etc.\@\xspace}%
}
\newcommand*{\etal}{%
    \@ifnextchar{.}%
        {et al}%
        {et al.\@\xspace}%
}
\makeatother

\newcommand{\fig}[1]{Figure~\ref{#1}}

\begin{document}

\title{Navigating Diverse Data Science Learning:\\Critical Reflections Towards Future Practice}

\author{\IEEEauthorblockN{Yehia Elkhatib}
\IEEEauthorblockA{\textit{MetaLab, School of Computing and Communications}\\
Lancaster University, United Kingdom\\
\{i.lastname\}@lancaster.ac.uk}
}

\maketitle
\thispagestyle{plain}
\pagestyle{plain}

\begin{abstract}
Data Science is currently a popular field of science attracting expertise from very diverse backgrounds. Current learning practices need to acknowledge this and adapt to it. This paper summarises some experiences relating to such learning approaches from teaching a postgraduate Data Science module, and draws some learned lessons that are of relevance to others teaching Data Science.
\end{abstract}


\section{Introduction}

As Data Science (DS) continues to be a growing field with promising prospects \cite{ibm,itjobs,pwc}, it is attracting significant attention from many including learners of different learning backgrounds and applications areas. 
From a DS educator's perspective, the result is a very diverse cohort of learners. This typically includes (in no order) mathematicians, statisticians, operations researchers, computer scientists of all their colours, other scientists (\eg environmental scientists, psychologists, \etc), and business managers and analysts to name a few. This in itself poses a number of challenges to the educator. In addition, as is common with emerging fields of science, teaching DS commonly starts as a specialist graduate-level course. In due course, it would establish itself as a stand alone undergraduate speciality. Until then, however, learners come in with a fairly dense knowledge, exposure to, and preference to certain learning approaches. This poses even further challenges to the educator. In fact, it also creates challenges to the learners themselves in terms of how to interact with colleagues who studied different disciplines, and are potentially accustomed to different learning methods and materials.

In this paper, I will focus on the challenges posed to the educator. I will use the term \textit{educator} to refer to the lecturer, professor, or teacher; and the term \textit{learner} to refer to the students or members of the learning cohort. 
I reflect on my own experiences of teaching on a MSc level course created specifically for DS. Established in 2014, the DS MSc course at Lancaster University was one of the first of its kind and continues to draw a large number of learners from all over the world. 
The cohort consists of 50-70 learners that primarily come from different backgrounds. 
The experiences related in this paper are drawn from a module I teach on this MSc course. The module is an introductory one intended to equip learners with basic data analysis and experimentation skills that are essential to industrial and academic DS work. The module covers a wide range of DS foundations in the span of 10 weeks. 

The paper provides the following contributions:
\begin{enumerate}
  \item Identify the make up of a DS practitioner and team structure.
  \item Explore some of the challenges associated with teaching DS (at a graduate level).
  \item Enumerate a number of successful and unsuccessful practices and approaches.
  \item Distill a number of lessons learned for the benefit of other educators in the DS field.
\end{enumerate}

\section{What's in a Data Scientist?}
\label{sec:what is}

Before delving into the challenges and experiences, it is important to ensure common understanding about what a data scientist is. While many others (\eg \cite{patil2011,conway2013,lyon2017education}) have focused on the skills of a `data scientist', I instead focus on their roles.

\subsection{Many Not One}
Some assume that a data scientist is a single, well-defined role. Indeed, it requires a unique set of skills that sets it apart from other established roles in modern ICT industry (\eg systems developer, or network engineer). However, as the DS market develops, so does our understanding of what it is capable of and entails. As such, I came to appreciate that a data scientist is neither a single role nor a necessarily well-defined one. Instead, it is a collection of different roles that complement each other. I separate these into \emph{core} and \emph{auxiliary} roles.

Before defining these, it is worth noting that the data scientist role is itself rather malleable and context specific. In one industry, for example, a data scientist could be someone who analyses data streams for insights that directly affect the tactical and strategic company directions, while in other industries the same term could be used to refer to someone who collects and curates data. Some of this divergence stems from misunderstanding of the different roles that a data scientist might have. And although these roles are varied but fluid, \ie distinction between them is not always clear even for the role holders themselves, understanding them helps bring some common understanding to what being a data scientist means. It also helps us to appreciate the diverse set of skills required, and better assemble the right teams and required support systems.

\subsection{Core Roles}
\begin{figure}[!ht]
	\centering
 	\includegraphics[width=0.8\columnwidth]{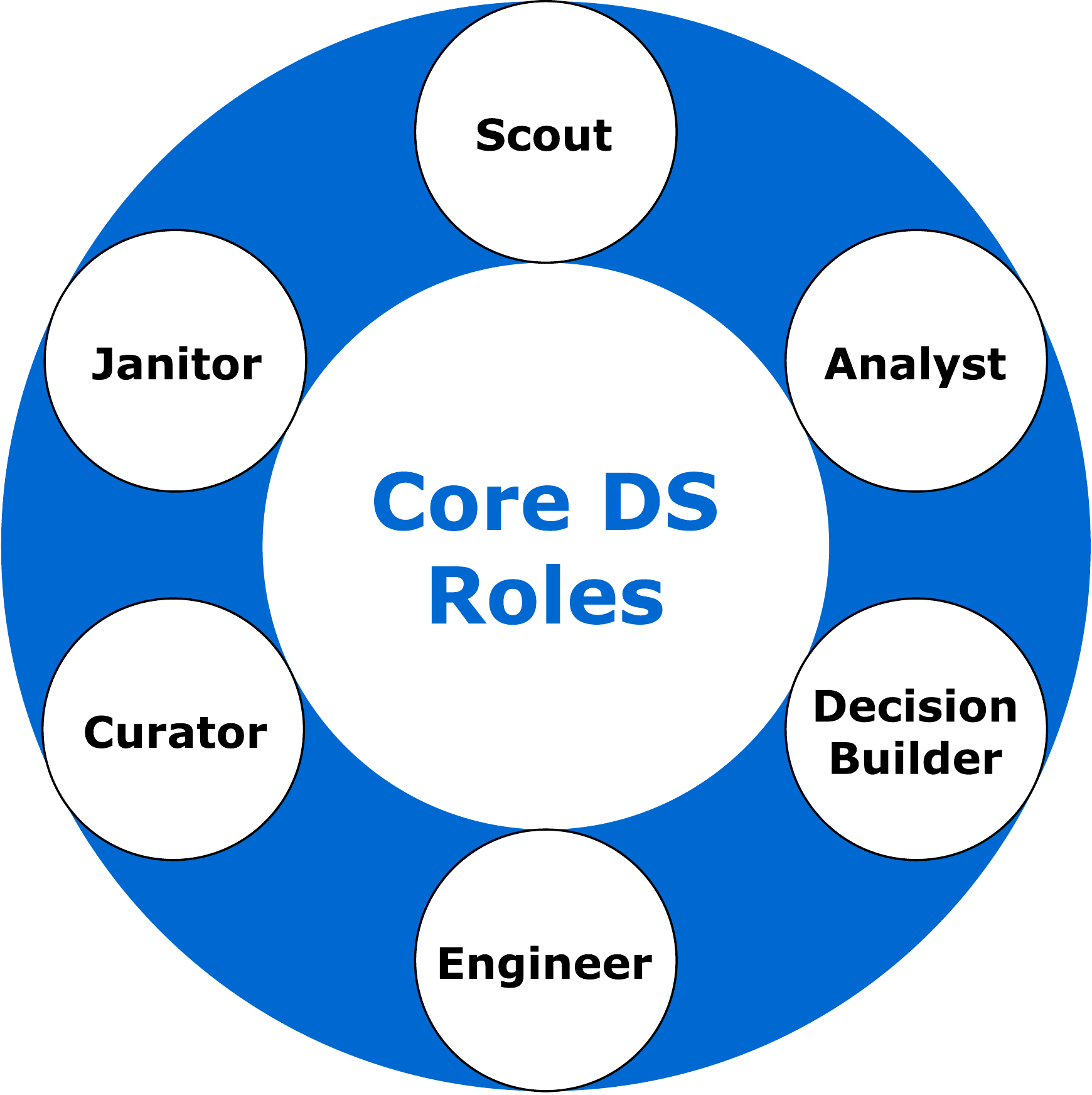}
	\caption{Core data scientist roles.}
	\label{roles-core}
\end{figure}

The core roles are described as follows and depicted in \fig{roles-core}.
\begin{itemize}
  \item \textbf{Janitor} is perhaps the role that is most hidden from view and is thus the most underappreciated. Data is rarely ever perfectly ready to use as is, and features a fair amount of missing data points, outliers, duplicates, and wrongly labelled data points. This is caused by capturing methods, sampling approach, or human/machine intervention on the path through which the data passed. Processing such inconsistencies manually comes naturally to many people especially those acquainted with the data provenance, but automated methods are not yet sophisticated enough to be able to reason about them in a completely unsupervised fashion. As such, the data janitor role entails a non-insignificant effort to clean and pre-process the data in order to prepare it for analysis.
  \item \textbf{Scout} performs exploratory data analysis for sanity checking and early insights. This uncovers data structure (if this is not known in advance) and identifies inconsistencies, both things that will help other roles that will work with the data at a deeper level. Also, the scout usually forms initial hypotheses that will seed others and feed into the work of other roles.
  \item \textbf{Analyst} is what most people attribute to being a data scientist. They dig deep into the data in order to extract meaning, discern  patterns, identify the essential chronicle of the data and what it describes, and uncover evidence of unforeseen narratives. This entails, at a high level, forming hypotheses and designing corresponding tests. The implementation of such tests could follow any of a number of methodologies.
  \item \textbf{Decision Builder} carries on the work of the analyst and builds products that will automate decision making or alternatively provide decision making support based on the outcomes of the analysis. This commonly includes adaptive machine learning and deep learning methods, with the aim of transforming the insights of the analysis into actionable decisions.
  \item \textbf{Curator} is responsible for holding and maintaining the data. This includes traversing concerns of storage formats, access interfaces, data governance, custodianship, and responsible sharing.
  \item \textbf{Engineer} defines different setups in order for other roles to be able to interact with the data efficiently and reliably. 
  They would also be responsible for managing the interface between development and production products and environments.
\end{itemize}

\vspace{-0.2em}
\noindent
Core roles sometimes overlap, and usually interact through iterative processes that need to adapt to changes in the incoming data and the analysis objectives.

\subsection{Auxiliary Roles}
Data scientists seldom work in isolation. They interact with teams responsible for creating data, they work with others who help them in their analysis, and they communicate with those who have a vested interest in the data. In fact, working in isolation renders their job meaningless beyond fascination with data.

\vspace{-0.2em}
\begin{figure}[!t]
	\centering
 	\includegraphics[width=0.8\columnwidth]{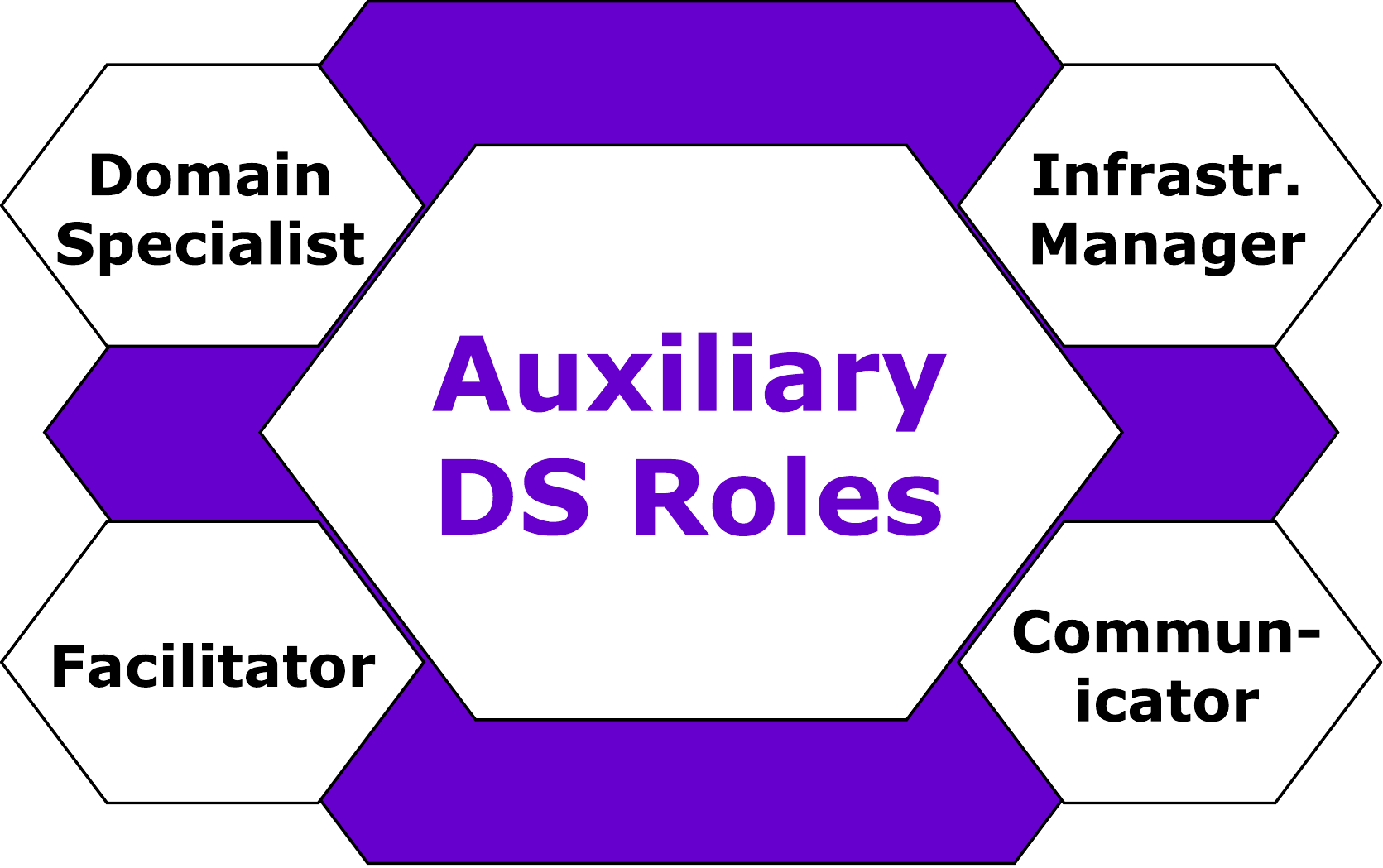}
	\caption{Auxiliary data scientist roles.}
	\label{roles-aux}
\end{figure}

This interaction along with the increasing sophistication of data science typically results in data scientists resolving to work in teams. 
Consequently, the distinction between the core roles described above begin to become clearer. Additionally, a number of other roles begin to emerge as DS teams grow. I refer to these as auxiliary roles (\fig{roles-aux}) and they are as follows.
\begin{itemize}
  \item \textbf{Domain Specialist} provides much needed domain expertise to help decipher provenance, data significance, sources of bias, and implications.
  \item \textbf{Infrastructure Manager} provides support to build and operate data systems beyond the role of the data engineer. For example, a data engineer might set up a Spark application pipeline and an associated development environment, whilst the systems developer would help streamline data pipeline production, and deal with the management of underlying system infrastructure.
  \item \textbf{Communicator} takes on the responsibility of communicating analytic outcomes outside of the visualisation of exploratory and confirmatory analysis results done by the core team. This includes building products for interacting with constructed data systems, interactive visualisation to communicate results to audiences outside the DS team (\eg business management), and creating easy-to-digest insights (\eg in the form of infographics).
  \item \textbf{Facilitator} provides additional support in terms of setting up systems to confirm or disspel certain hypotheses that are emerging from the core team; \eg setting up and carrying out A/B experiments, or procuring external data sets.
\end{itemize}


\section{Curriculum Design or Implementation?}
\label{sec:delivery}

There is a real need to not only focus on the learning outcomes and approaches, but also on the practical means of implementing these approaches and to identify the exact structure by which the learning outcomes are to be achieved. This is something that in many ways transcends the design of a course; it drills deep into how each pedagogical element is being created and delivered, and needs continuous monitoring.

Here, I will focus on in-class activities as an example. I put emphasis on discussion in class, giving learners between 1 and 3 group learning activities during every lecture, usually ending with a few minutes of open discussion. Instead of parlaying knowledge via PowerPoint, this method encourages continuous stimulation of the learners' critical and creative thinking skills through direct fundamental questions, group brainstorming, rhetorical questions, application to top tier papers, and coming up with solutions to practical dilemmas \cite{McLean1997,Steinhert1999,Shulman2005,Hardin2015think}. They also encourage learners to know each other, and for the educator to know the learners, and additionally to gauge comprehension and application across topics and subjects. (More on this in \S\ref{sec:learn}.)

One thing that soon became noticeable is the difference between the performances of learners with a CS background compared to those of other backgrounds. This was unexpected as in-class activities were mainly about applying concepts and exercising analytical skills followed by a discussion, and never included any programming skills of any kind. Upon closer inspection, I realised that the difference was in fact between home and overseas learners. 

I discussed this informally with a small subset of overseas learners. It soon became apparent that this issue relates directly to the learning setting, which is unfamiliar to some due to their previous experiences. Some overseas learners found it difficult to participate in group exercises, especially when they had doubts about their understanding of the lecture material. Home learners, on the other hand, were generally much more confident in articulating their understanding (regardless of how correct it may be) and expressing their views, which made them undermine their own and shy from in-class participation. It is important to note that this is \textsl{not} due to a language barrier as all learners are comfortable using and comprehending English at postgraduate level.

I was able discern two parts to this issue. The first relates to the input: some, if not many, overseas learners are not used to being asked to apply critical thinking to what the educator tells them, or what they read in a book or an academic paper. This creates a barrier to applying their knowledge and also to gain new knowledge through application and discussion. The second part of this problem relates to the output: many, especially overseas, learners fail to realise that learning is largely a social activity \cite{Orsmond1997,Qadir2015}. Perhaps they were (either in their previous educational institution or discipline) not encouraged to share with their colleagues, possibly even implicitly trained to deal with colleagues as competitors. This instils an unwillingness to participate in group activities and thus learners miss out on one of the key learning activities of the module. 


My approach to solving these two issues was simple and effective. First, I made clear and explicit remarks about how \textbf{no work is infallible}, including the ones I introduce as part of their learning activities. Before each group exercise, I gave brief examples of how to critique similar work and ways of improving said work. Second, I made conscious effort to interleave my lecture material with \textbf{checkpoints}; these are frequent but short pauses where I very briefly reflect with the learners on something I just introduced to or discussed with them, allowing them a moment to focus on the processes and not just the artefacts / outcomes \cite{Rowe2015}. Third, I designed clear guidelines to what I expect (and, perhaps more importantly, do not!) each group discussion to produce, and I made these \textbf{expectations} as minimal and elemental as possible. My rationale was that any learner can easily apply her/ his thinking to essentially ``fill in the blanks'', encouraging them to participate through a low barrier whilst explaining their reasoning in a step-by-step manner \cite{ALEVEN2002}.

As a result learner participation was very rapidly increasing, no longer with any clear distinction between home and overseas learner participation input. This was also reflected in their coursework submissions.

\section{Contextualise This}
\label{sec:context}

There are a number of well known factors that help effectively attain learning outcomes such as use of clear language, adequate level of complexity given the cohort's educational attainment, appropriate information given lecture and module duration, clear setting of expectations and learning outcomes, and appropriate use of examples. However, a common but hidden thread through these is contextualisation, which is well documented (\cf \cite{cordova1996,holbrook2010,HOOD201583}) to allow learners to use their own set of skills and understanding, to appreciate the relevance of the learning material, and consequently to be motivated to actively engage with the course material.

Let us focus on the use of examples to discuss this. I tend to use real world examples whenever possible in order to help contextualise the learning material and make them relatable as much as possible. 
For instance, when describing what data bias is, I give plenty of examples from industry and academia detailing different techniques to quantify and identify biases, and how to set up processes to identify potential bias and its effect on validity. 

Based on clear feedback from the learners, this form of contextualisation helps them tie new concepts with old ones, and to illuminate routes from theory to practice. However, it is also not easy. It takes a lot of time to think of appropriate examples that the learners would be able to relate to both at the time of the lecture and when they move beyond the course. This is especially true when considering their various backgrounds. 

There is no silver bullet here. My approach is to match the learners' backgrounds and use examples from as varied DS-related fields as possible. In other words, rather than restricting to examples from my own research areas of distributed computing and networking \cite{Elkhatib2014CanSPDY,Elkhatib2014browsing,Blair2015holons,elkhatib2017microclouds}, I \textbf{actively seek examples from further fields} through engaging with colleagues from other CS sub-disciplines as well as other disciplines like environmental science, politics, psychology, accounting and finance, \etc

\section{Assessment Rationale and Analysis}
\label{sec:assessment}

The coursework structure relied on different elements: 
\emph{knowledge}-based elements to test the grasp of information (in the form of simple in-class questions, as described above), 
\emph{comprehension}-based elements to gauge the understanding of information, and 
\emph{application}-based elements to evaluate problem solving abilities. 
This section focuses on the latter two elements and their associated challenges.



\subsection{Comprehension-based Assessment}
\label{sec:assessment:c}

Knowledge-based assessment take the form of straight-forward questions. These were embedded in the lecture delivery strategy during regular checkpoints, as explained in \S\ref{sec:delivery}. 
Modifying these into comprehension exercises is important in order to raise discussion and engender deeper understanding. 
Self-assessment is an established technique where learners appraise their own understanding with minimal guidance \cite{Orsmond1997}. 
I relied on \textbf{mutual-assessment}, a slightly modified version of self-assessment where learners appraise each other's understanding. This was set up as follows. 
Once or twice during each lecture, the learners are given a problem that would test their understanding of the lecture material thus far. They are asked to write their answers on paper within 3-5 minutes, then swap the answer sheets with one of their colleagues.\footnote{They could swap with their nearest neighbour if time is constrained. Otherwise, I ask them to move around the room.} 
They would then give each other feedback about how they fared, and discuss for another 3-5 minutes. The whole class would then reconvene for another 3-5 minutes to have a wider discussion of the main points, or sometimes to answer questions about things the learners could not clarify to each other. Many chose to share this verbally with the rest of the class, whilst some preferred to write it down on provided notes that I collect and read out. 

This approach was extremely successful for four main reasons. First, the learners were able to test and enhance each other's understanding, and provide very personalised feedback to one another (much more than I could due to the scale of the class). This is even more effective than in-class interactive and online quizzes, which I tried in an earlier module, where feedback was inevitably brief, and too factual despite being personalised. 
Second, this approach is very practical as it scales really well, reducing the amount of individual marking the educator has to do and allows them to focus on understanding their cohort.  
Third, the conclusion of each exercise provides the educator with a vital feedback loop. In effect, it signals the parts of the learning material that are the `muddiest point' \cite{mosteller1989} that might have been explained better, and whether these were common to many learners or restricted to only a subset. Moreover, due to the huge diversity of this particular cohort, the feedback is generally quite wide-ranging and many times includes things I myself would not have thought of. This was hugely educational and eye opening learning approach. 
Fourth, and this leads to the topic of the next subsection, is the promotion of a lively classroom culture where there is close interaction between members of a cohort that would otherwise fall into cloistered cliques of CS, DS, \etc 

One complication, though, is the reliability and fairness of the marks provided by the learners. To ensure this, I needed to devise an additional screening process to moderate the marks given by the learners. This, however, is not a great burden and is rather manageable and reasonable compared to the gained benefits. Reflecting on this, a potential future direction of improvement is to set a loose marking scheme for the learners to use as a reference when marking each others' work.

\subsection{Application-based Assessment}
\label{sec:assessment:a}

As highlighted in \S\ref{sec:what is}, there are different roles that any data scientist might take. Common practice in industry is to assemble DS teams where different data scientists would take on one or more of these roles and work closely together \cite{Dinter2015}. Accordingly, application-based assessment elements are structured around \textbf{group work} where learners work together to tackle a certain data challenge. There are a number of useful lessons in this regard.

First, the data problems that the learners are asked to tackle are all real ones provided by industrial partners. As such, they give a real sense of the challenges that DS teams are currently facing in industry. This is very appealing to learners and helps in maintaining their engagement with the required work even in the face of difficulties. In fact, many of our learners continue to work with the industrial partners beyond the course on very similar challenges. The main restraint here relates to drawing important challenges from industrial partners and to involve them in defining the student projects without expecting a significant time commitment from their side.

Second, the way learner teams are set is of crucial importance. If they set the teams themselves, the teams tend to be exhibit high self-similarity in terms of background and skills. If, instead, they are set up by the educator, there is a risk of coercing incompatible personalities to work together. Hence, a hybrid approach was adopted: At the beginning of term, learners are asked to express their skills and experiences through self-assessment forms. The form includes Likert scale scores on different abilities such as \textit{`Statistical Modelling'} and \textit{`Data Handling'}. The scores are then used as input to a clustering algorithm that produces teams of 4-5 learners with a good balance of skills across each team. The produced teams are then proposed to the students, giving them an opportunity to gauge whether they would be interested to work with their classmates. After such stage, many learners accept the created teams or slightly adjust them. As an added bonus to this exercise, the self-scores are revisited during the last lecture to give learners the opportunity to reflect on how their abilities and skills progressed through the module. Incidentally, several learners have observed how this was useful further still by helping them be mindful of their perceived knowledge.

Third, the ability to work well within a team is not something that is initially attained by all learners. Also, learners of different backgrounds rely on different collaborative systems. An explicit point is made to the learners that teamwork is a crucial part of their training in light of the nature of DS teams (see \S\ref{sec:what is}). General advice is given to the learners about how to allocate and monitor group work. Additional guidance is also provided on both individual and group levels. Plenty of practical tips are supplied regarding things that are not covered by the syllabus such as collaboration and peer review tools (\eg GitHub, Jupyter~\cite{4160251}, and other generic \cite{tennant2017review} and field-specific tools \cite{Vitolo2015web,Greene2015}) and best practices regarding project management and presentation strategies that are suitable for academia and industry. 

Finally, assessment criteria needed to be more detached from the means of accomplishment. In other words, the application-based assessment needs to be able to accommodate a wide range of preferences in terms of processing tools, programming languages and frameworks, and presentation styles. For instance, the learners are given flexibility to accomplish the group work using Python, Java, R, or Matlab as long as the end results are presentable in a format that is suitable for an academic or managerial audience, and the code and artefacts are clearly annotated and self-describing. Obviously, this raises the expectations on the educator but is a normal state of affairs for any application-based course, especially considering the diversity of DS roles and learner backgrounds. 

%

\section{Learning the Learners}
\label{sec:learn}

Beside the aforementioned strategies and practices, the educator needs to realise the importance of knowing their learners well. There is a great deal to be learned from teaching literature, peer observation, and critical evaluation approach. However, the actions needed to implement these best practices and methodologies all seem to hinge on simply getting to know the learners in terms of their backgrounds, abilities, interests and experiences, and to consequently act on this recognition to provide a \textbf{tailored learning experience}~\cite{Manning2010}. 
For example, I design my lectures to leave plenty of room for self exploration and creative thought, assembling a myriad of avenues of inquisition in different directions. However, on deep reflection I came to realise that Socratic teaching methods are not enough. Extra work is required in order to cater to learners with diverse learning backgrounds and skills. 

Realising this requires looking up from the lecture material, knowing the learners through direct interactions, and tuning the fine details of the syllabus to best suit the learners' abilities and experiences. Such fine details include, for instance, adopting an in-lecture checkpointing practice (\S\ref{sec:delivery}) and allowing learners to reflect on the subject, assisting them in dissecting it and exposing the strengths and weaknesses (\S\ref{sec:assessment:c}). Another example where knowing learner abilities is important is when contextualising lecture material (\S\ref{sec:context}), and planning group efforts (\S\ref{sec:assessment:a}).

This level of knowing the learners is not something that is inherently included in a course syllabus, nor is it something that can be easily allocated as a time-limited task (such as delivering a lecture, supervising a lab, \etc). Instead, it is an underlying activity that an educator is responsible for in order for them to ensure optimal module delivery. Furthermore, this is not necessarily something that requires a significant amount of effort or one that might incur big changes to curriculum. This is easily blended into the learning approach through lectures, lab sessions, and assessment exercises as demonstrated through examples in this paper. 
The key is to introduce cognisance for a tailored delivery that is easy to track and tweak, if and where necessary.

\section{Final remarks}



Any educator working with a diverse cohort of learners stands to greatly benefit from \textbf{observing practices in other fields of science}. Due to the interdisciplinary nature of our DS course at Lancaster, I had the opportunity to interact with colleagues from across the university and assimilate their educational approaches. Although starting points and goals are quite similar in a pedagogical sense, the methods are often quite different. I attribute some of this to disciplinary differences and conventions, but also to variances in learning backgrounds and previous experiences. This forced me to reflect on my own practices in a critical light, and also to identify the distinctions between approaches and recognise their development. Learners are different and thus educators need to expand their educational toolboxes to cater to such diverse cohorts.


\balance{
	\bibliographystyle{IEEEtran}
	\bibliography{bibs}
}

\end{document}